\documentclass[10pt,nocopyrightspace]{sigplanconf}
\usepackage[usenames,dvipsnames]{color}

\usepackage{mathpartir}
\usepackage{amsthm}
\usepackage{stmaryrd}
\usepackage{amssymb}

\usepackage{listings}

\usepackage{pstricks, pst-node, pst-plot, pst-tree}

\usepackage{booktabs}

\usepackage[colorinlistoftodos]{todonotes}

\usepackage{url}

\lstdefinelanguage{JavaScript}{
        keywords={      attributes, class, classend, do, empty, endif, endwhile, fail,
                                function, functionend, if, implements, in, inherit, inout, not, of,
                                operations, out, return, set, then, types, while, use, else, switch, case,
                                break, default, for, var},
        keywordstyle=\color{blue}\bfseries,
        ndkeywords={trace, permit, apply, applyObj, permitArgs, applyArgs},
        ndkeywordstyle=\color{blue}\bfseries,
        identifierstyle=\color{black},
        sensitive=false,
        comment=[l]{//},
        morecomment = [s]{/*}{*/},
        morecomment = [s][\color{green}]{/**}{*/},
        commentstyle=\color{gray}\ttfamily,
        stringstyle=\color{red}\ttfamily
}
\usepackage{amsmath}

\begin{document}

\setlength{\pdfpageheight}{\paperheight}
\setlength{\pdfpagewidth}{\paperwidth}

%\conferenceinfo{DLS~'13}{October 28, 2013, Indianapolis, Indiana, USA}
%\copyrightyear{2013} 
%\copyrightdata{978-1-4503-2433-5/13/10}
%\doi{2508168.2508176} 

% Uncomment one of the following two, if you are not going for the 
% traditional copyright transfer agreement.

%\exclusivelicense                % ACM gets exclusive license to publish, 
                                  % you retain copyright

%\permissiontopublish             % ACM gets nonexclusive license to publish
                                  % (paid open-access papers, 
                                  % short abstracts)

\titlebanner{}        % These are ignored unless
\preprintfooter{On the Proxy Identity Crisis}   % 'preprint' option specified.

%\title{On the Equality of JavaScript Proxies}
%%\title{Identity vs. Equality for Proxies}
%%\subtitle{Full Presentation}

%\title{How (un-)equal is equal?}
%\subtitle{Identity vs. Equality for Proxies}

%\title{Identity, Equality, Fraternity}
\title{On the Proxy Identity Crisis}

\authorinfo{Matthias Keil \and Peter Thiemann}
{
%Institute for Computer Science\\
University of Freiburg,
Germany
}
{\{keilr,thiemann\}@informatik.uni-freiburg.de\vspace{-\baselineskip}}

\maketitle

%\input{abstract}

%\category{D.3.3}{PROGRAMMING LANGUAGES}{Language Constructs and Features}[Classes and objects]
%\category{D.3.3}{SOFTWARE ENGINEERING}{Software/Program Verification}[Programming by contract,Validation]
%\category{D.4.6 }{OPERATING SYSTEMS}{Security and Protection}[Access controls]

% general terms are not compulsory anymore, 
% you may leave them out
%\terms
%Design, Languages, Security, Verification

%\keywords
%Access Permission Contracts, JavaScript, Proxies

\section{Introduction}

A proxy, or wrapper, is an object that mediates access to an arbitrary target object.
%The objective of introducing a proxy is to extend or restrict the functionality of the underlying object. 
Proxies are widely used to  perform resource management, access remote objects, impose access control \cite{VanCutsem:2010:PDP:1869631.1869638,Keil:2013:EDA:2508168.2508176}, restrict the functionality of an object \cite{DBLP:conf/oopsla/StricklandTFF12}, or to  enhance the interface of an object. %XTend, Expanders \cite{DBLP:conf/oopsla/WarthSM06}
Ideally, a proxy is not distinguishable from other objects so that  running a program with an interposed proxy should lead to the same outcome as running the program with the target object, unless the proxy imposes restrictions.

Proxies introduce a subtle problem. Because a target object may have any number of proxy objects, which are all different from the target, a single target object may obtain multiple identities---it suffers from schizophrenia!
Even worse, it turns out that there is no single cure for this schizophrenia because the desired behavior depends on the use  case. 

Unfortunately, current proxy implementations are committed to particular use cases, which makes it hard to adapt them to  uses with different requirements.  We discuss two such use cases in the context of the JavaScript proxy API  \cite{VanCutsem:2010:PDP:1869631.1869638}, identify its shortcomings, and propose a solution. 
% \cite{ van2012design}

\subsection{JavaScript Proxies}

The JavaScript proxy API \cite{VanCutsem:2010:PDP:1869631.1869638} provides a proxy constructor that takes the proxy's target object and a handler object:
\begin{lstlisting}
var p = new Proxy (target, handler);
\end{lstlisting}
The handler object provides optional trap methods that are invoked when operations are applied to the proxy.
For example, a property get like \texttt{p.foo} invokes the trap \texttt{handler.get(target,'foo',p)} if that trap is present. 
%% RMK: remove figure because of space reasons
%%Figure~\ref{fig:prxie_pattern} illustrates this scenario, where the handler \texttt{h} forwards the operations on the proxy \texttt{p} to the target object \texttt{t}.
Untrapped operations are forwarded to the \texttt{target} object.

The JavaScript proxy API treats proxies as \emph{opaque}: each proxy object has its own identity different from all other (proxy) objects and this difference is observable with the JavaScript equality operators  \texttt{==} and \texttt{===}. When applied to two objects, both operators compare the object references.\footnote{If one argument has a primitive type, \texttt{==} attempts to convert the other argument to the same primitive type, whereas \texttt{===} returns false if the types are different. If both arguments are objects, then both operators do the same.}
%% RMK: remove figure because of space reasons
%\begin{lstlisting}[float,caption={Comparing opaque proxies.},label={lst:comparing-proxies},captionpos=b]
%var obj = {a: 5, b: 42};
%var p = new Proxy(obj, {});
%var q = new Proxy(obj, {});
%var eq1 = (p==q); // false
%var eq2 = (obj==p); // false
%\end{lstlisting}
%Listing~\ref{lst:comparing-proxies} demonstrates that comparing distinct proxies returns false even though the underlying target is the same. Similarly, an unwrapped target object  is not equal to any of its proxies.
The use of equality has one consequence: comparing distinct proxies returns false even though the underlying target is the same. Similarly, an unwrapped target object  is not equal to any of its proxies.

\subsection{Use Case: Access Control}
\label{sec:use-case-access-control}

JavaScript proxies implement access control wrappers like revocable references and membranes in a library \cite{VanCutsem:2010:PDP:1869631.1869638}. The idea of a revocable reference is to only ever pass a proxy to an  untrusted piece of code, e.g., a mashup. Once the host application deems that the mashup has finished its job, it revokes the reference which detaches the proxy from its target. Membranes extend this method recursively to all objects reachable from the object passed to a mashup. Opaque proxies are required for implementing this library.

The JavaScript proxy API is tailored to uses where access is strictly compartmentalized. The host application  only sees the original objects whereas the mashup only sees proxies. Furthermore, the implementation of revocable references and membranes ensures that there is at most one proxy for each original object. For this reason, each compartment has a consistent view where object references are unique.

\subsection{Use Case: Contracts}
\label{sec:use-case-contracts}

Proxies implement contracts in Racket \cite{DBLP:conf/oopsla/StricklandTFF12} and in JavaScript \cite{Disney2011,Keil:2013:EDA:2508168.2508176}. Contracts impose restrictions that the programmer regards as preconditions for the correct execution of a program. For example, a contract may require a method to be called with a particular type or an object property to always contain positive numbers.

During maintenance, the programmer may add contracts to a program as understanding improves. Clearly,  the addition of a new contract must not change a program execution that respects the contract already. In this scenario, the program executes in a mix of original objects and proxy objects. Furthermore, there may be more than one proxy (implementing different contracts) for the same target. If introducing proxies affected the object identity, then some true comparisons would flip to false, thus changing the semantics.

Consequently, the Racket implementation provides \emph{transparent} proxies \cite{DBLP:conf/oopsla/StricklandTFF12}, which are indistinguishable from their target object, recursively. 
%% RMK: remove figure because of space reasons
%Listing~\ref{lst:comparing-transparent-proxies} shows the desired behavior of hypothetical transparent proxies in JavaScript.
%\begin{lstlisting}[float,caption={Desired behavior of transparent proxies.},label={lst:comparing-transparent-proxies},captionpos=b]
%var t_p = new TransparentProxy(obj, {});
%var t_q = new TransparentProxy(obj, {});
%var t_r = new TransparentProxy(t_q, {});
%var t_eq1 = (t_p==t_q); // true
%var t_eq2 = (obj==t_p); // true
%var t_eq3 = (obj==t_r && t_p == t_r); // true
%\end{lstlisting}

\subsection{Assessment}
\label{sec:assessment}

Neither the opaque nor the transparent proxy implementation can be labeled as right or wrong without further qualification.
Each is appropriate for a particular use case and leads to undesirable behavior in another use case.

It is also clear that the behavior of equality is not something the should be left to the whim of the programmer. For example, equality on objects should be an equivalence relation, which means that the equality operations \texttt{==} and \texttt{===} must not be trapped \cite{DBLP:conf/ecoop/CutsemM13}.

Thus, the current state of affairs in JavaScript is fully justified, but it is not well suited to implement contract systems.  Hence, we explore some alternative designs that would suit both use cases.

\section{Alternative Designs}

%\section{A new equals operator}
\paragraph{Proxy-aware equality}

%Unfortunately, there is no easy way to address this shortcoming.

One way to obtain transparent proxies with an implementation of opaque proxies
is to provide proxy-aware equality functions like \mbox{\texttt{Proxy.isEqual()}} and \mbox{\texttt{Proxy.isIdentical()}} to replace all uses of \texttt{==} and \texttt{===}, respectively, in an application program. 
%
%One possible solution would be introduce a quadruple equality operator \texttt{====} extending the triple equality operator by possibility to see through a proxy. 
%
% \begin{lstlisting}
% var p = new Proxy(obj, {});
% var q = new Proxy(obj, {});
% var eq1 = (p==q); // false
% var eq2 = Proxy.isEqual(p,q); // true
% \end{lstlisting}
%
This approach preserves the previous behavior and retains the possibility to distinguish proxies from target objects in library code implementing proxy abstractions. However, it would require the application code to be transformed (at run time to support \texttt{eval}), which is not feasible in an application like access control \cite{Keil:2013:EDA:2508168.2508176} that must work with unmodified foreign code.

\paragraph{Transparent Proxies}

Making proxies generally transparent makes it impossible to test whether a reference is a proxy or an original object. 
However, there are abstractions that require such a test. For example, our implementation of access permissions \cite{Keil:2013:EDA:2508168.2508176} extracts the current permission from a proxy to construct a new proxy with an updated permission. This improves the efficiency of the implementation, which would otherwise generate long chains of proxy objects. 

Thus, for implementing proxy abstractions it must be possible to break the transparency.

\paragraph{More equality operators}

Another possible solution would be to reinterpret the JavaScript equality operators \texttt{==} and \texttt{===} as proxy-transparent
%% RMK: remove figure because of space reasons
%(as in Listing~\ref{lst:comparing-transparent-proxies})
and introduce new variants, say, \texttt{:==:} and \texttt{:===:} for their opaque cousins. 
The former operators are supposed to be used in application code whereas the implementation of proxy abstractions could make use of the opaque operators where needed. 

No code transformation is required with this approach. However, it is not clear how to ensure that application code does not use the opaque operators. It is not even clear if it \emph{should not} use them. While proxy abstractions can be implemented, the distinction between application and library seems too rigid.
Given both operations, application code can test if one object is a proxy for another:
\begin{lstlisting}
var isProxy = ((objA==objB) != (objA:==:objB));
\end{lstlisting}

%transparent 
%Proxies could be extended to be transparent with respect to the double and triple operator.  
% Opposed direction as bevore

% \begin{lstlisting}
% var p = new Proxy(obj, {});
% var q = new Proxy(obj, {});
% var eq1 = (p==q); // true
% var eq2 = (p====q); // false
% \end{lstlisting}

% A developer gets the ability to distinguish proxies as needed. A code transformation is not required. 

%%%% 

\paragraph{Trapping the equality operation}

We already discussed that trapping the equality operation is not appropriate. However, there is a twist that enables modifying the equality without destroying its properties. Essentially, the handler is extended with a boolean trap:
\begin{verbatim}
isTransparent : function () -> boolean
\end{verbatim}
If the handler's trap returns false or if it is not present, the associated proxy behaves opaquely, otherwise it behaves transparently.  
The implementation is an extension of the equality comparison in the VM. 
Before testing reference identity as the last step in a comparison of two objects, 
the equality comparison calls a new internal \texttt{GetEqualityObject} method. 
For a standard object, this method returns its receiver.
For a proxy object, if is \texttt{isTransparent()} on the handler returns false, then \texttt{GetEqualityObject} returns the  reference to the current object.
Otherwise, it recursively invokes \texttt{GetEqualityObject} on the proxy's target.
For consistency, the \texttt{GetEqualityObject} method also needs to be called in other computations that depend on object identity, for instance the WeakMap abstraction provided by some JavaScript implementations.

This design enables both scenarios described in Sections~\ref{sec:use-case-access-control} and~\ref{sec:use-case-contracts} by configuring the handler appropriately. It also guarantees that equality is an equivalence relation in application code that does not have access to the handlers. To implement proxy-based abstractions, it may be necessary to temporarily make proxies opaque. But opaqueness can be obtained by reconfiguring the handler in the library code, analogous to the  implementation of revocable references. To maintain consistency at the application level, it may be necessary to restrict modifications to this configuration to a certain scope analogously to dynamic variables \cite{DBLP:conf/pldi/HansonP01}.

% This approach is 
% VM's can implement this efficiently. 

\section{Conclusion}

We have shown that neither the transparent nor the opaque implementation of proxies is appropriate for all use cases.
We discuss several amendments and propose a flexible solution that enables applications requiring transparence as well as opacity. We are currently implementing this solution in a JavaScript VM and expect to report results soon.

% We have seen, there is no easy way to avoid the limitations of using proxies. Neither the unique proxy nor proxy-aware equality functions would solve the problem completely. The best solution would be to trap the equals operation. This approach guarantees indistinguishability on the one hand, and a user-defined distinguishability on the other hand. Trapping the equality would also solve the problem with comparing the proxy with its target.

% This solution would require some rewriting of the proxy API and of the existing comparative operators, which is less expensive than rewriting existing source code. 

% tehoretical other solutions,
% static flag
% oder simply changing the equals operator

\bibliographystyle{abbrvnat}

%\bibliography{newbib,bibliography,bibliography-tbda,abbrevs,papers,collections,theses,misc}
\bibliography{newbib,bibliography}

\end{document}